# Illuminating extracellular vesicles biology with super resolution microscopy: Insights into morphology and composition

Lucile Alexandre*[1], Daniele D'Arrigo*[1,2], Nicolas Kuszla[1], André Cronemberger Andrade[1], Amanda K A Silva[1], Cataldo Schietroma[2], Florence Gazeau[1], Quentin Lubart[2], Stéphanie Mangenot[1]

1- Laboratoire NABI NAnomédecine, Biologie extracellulaire, Intégratome et Innovations, CNRS UMR8175, INSERMU1334, Université Paris Cité, 45 Rue des St Pères, 75006 Paris, France
2- Abbelight, 191 Avenue Aristide Briand, 94230 Cachan, France

## Abstract

Super-Resolution Microscopy (SRM) is emerging as a powerful and innovative tool for imaging, characterizing, and understanding the structure of Extracellular Vesicles (EVs). By addressing the need for single-particle analysis with the high resolution required to study the composition and organization of these nanoparticles, SRM provides unique insights into EV biology. However, its application is accompanied by significant challenges, ranging from experimental setup to data analysis. This review outlines the fundamentals of SRM and its position within the broader field of EV research. We then explore its applications in evaluating (i) the morphological structure of EVs, (ii) their molecular composition, and (iii) their roles in biological systems. By offering practical guidance and an overview of critical parameters for standardization, this review aims at providing researchers with the tools and insights necessary to effectively apply SRM to EV investigation.

**Introduction**

Extracellular Vesicles (EVs) are nanometric lipid bilayer vesicles released by cells (Figure 1.A), essential for cell-to-cell communication by transferring proteins, lipids, and nucleic acids[1]. They are promising biomarkers for various diseases, as they can be found in a wide range of biological fluids, as shown in Figure 1.B. They play a key role in regenerative medicine, and offer intriguing potential for drug delivery solutions. EV research is rapidly evolving, and it is gaining relevance both in applied biology and in clinics. The term EV typically refers to all lipid bilayer-enclosed particles derived from cells, with size ranging from 25 to more than 1000 nm (Figure 1.C)[2,3]. The size range of EVs varies significantly depending on the cell origin, stimulating trigger and isolation technique employed in the experimental protocol. It has been demonstrated that, besides the size, the EVs consist in a highly heterogeneous mixture of different subpopulations that can be discriminated according to both biological (biogenesis, markers, cargo, in yellow in Figure 1.D) and physical (morphology, density, surface charge, biomechanical, in blue in Figure 1.D) properties[4,5]. This heterogeneity, together with the small size of the EVs, makes their isolation and characterization challenging[6].

An elective methodological approach for EVs isolation is still missing and usual methods often result in the identification of distinct EV subpopulations[7]. Affinity isolation methods are growing in use for EV purification but are limited by the lack of a universal EV marker. Centrifugal approaches remain the gold standard for EV purification. They are based on the size and density (ultracentrifugation, differential centrifugation, density gradient or cushion centrifugation) of the EVs. Filtration and chromatographic techniques rely mostly on the size (ultrafiltration, Size Exclusion Chromatography SEC) but can also be based on different physical (surface charge) or biological (immunoaffinity) features. Flow-based approaches exploit the differences in geometric size (deterministic lateral displacement, tangential flow fractionation), hydrodynamic size (Asymmetrical Flow Field Flow Fractionation AF4), surface charge (dielectrophoresis, electrical AF4) or other specific properties to isolate EVs and separate different subpopulations. Innovative solutions (acoustic separation, polymer precipitation) are continually being developed to improve EV isolation. Nevertheless, EV size remains one of the most critical parameters in both preparation and characterization processes.

Understanding the differences between EV subpopulations is crucial both for their potential use as biomarkers and for their engineering as nano-delivery vehicles[8,9]. These differences influence EV content and modulate interactions with recipient cells, ultimately leading to diverse effects on cellular behavior. EVs play a pivotal role in cell-to-cell communication, regulating various pathophysiological processes such as the immune response, immunomodulation, tissue regeneration, cancer progression, neurodegenerative diseases, and inflammation[10–13], relying on a wide range of bioactive molecules (proteins, nucleic acids, lipids and other biomolecules) loaded within and on the EVs. This molecular cargo varies according to the type and the pathophysiological conditions of the secreting cell. While separating EV subpopulations based on their biogenesis could provide valuable insights into their clinical relevance, it remains a significant technical challenge due to overlapping physical and molecular characteristics[14]. As a result, many studies instead focus on isolating EVs based on size and density[15–17]. Size-based separation has gained substantial attention because EV subtypes differ not only in their biophysical properties but also in their biological functions[4] (Figure 1.C). However, size-based isolation does not necessarily align with distinctions based on biogenesis. This highlights the need for precise characterization of EVs and of their biological and physical features.

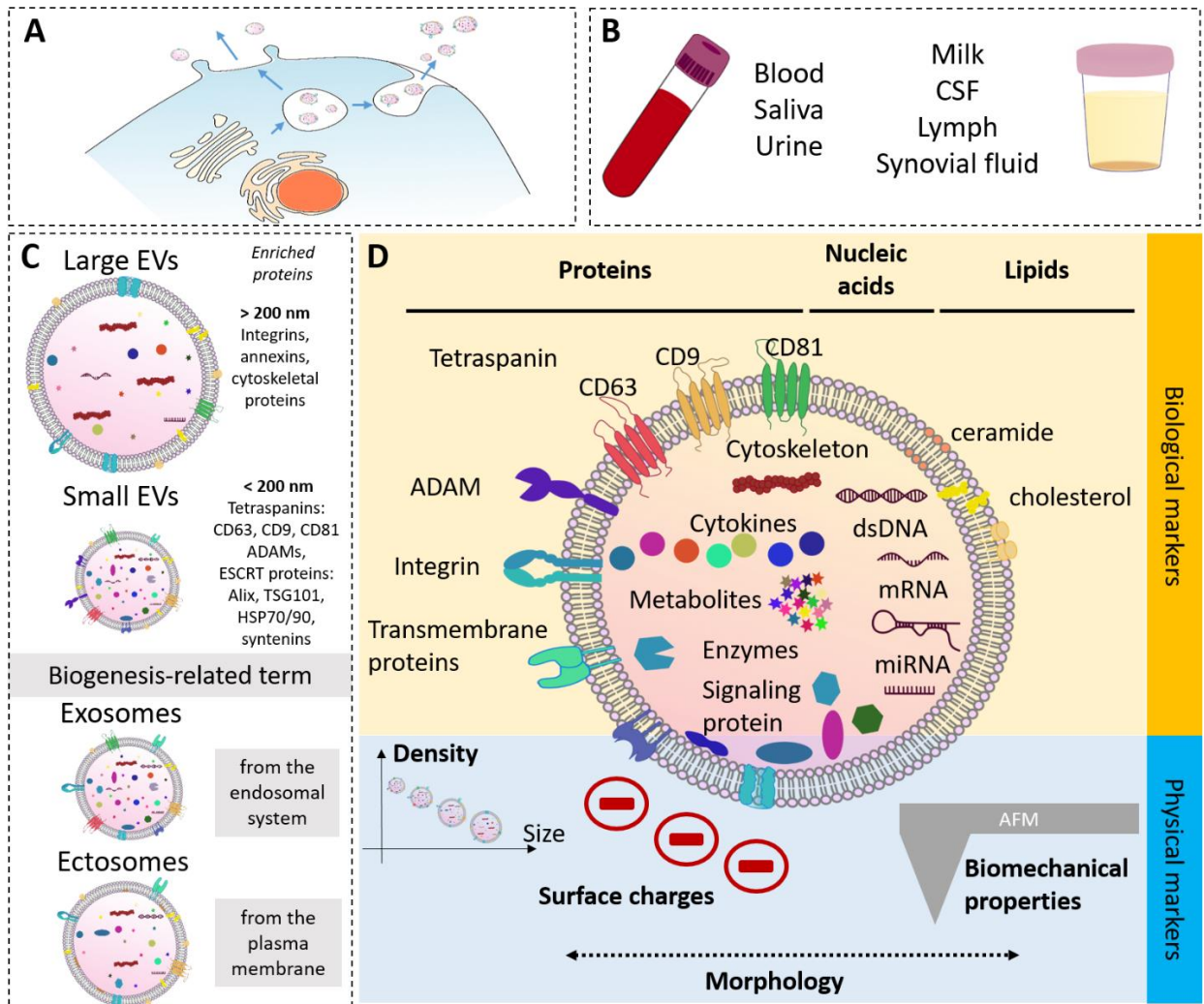

*Figure 1: Extracellular vesicles in biological samples A) EVs are produced by cells either through direct budding from the plasma membrane or via exocytosis of multivesicular bodies, blue arrows show the release paths B) EVs are present in different biological fluids. C) EVs can be separated into several categories depending on their size: large EVs (diameter larger than 200 nm) and small EVs (diameter smaller than 200 nm) or depending on their biogenesis: exosomes (originated from the endosomal system)*

*and ectosomes (originated from the plasma membrane). For small and large EVs, overexpressed proteins have been associated to each sub-population (right column). D) The division in sub-populations relies on the presence of biomarkers, in the form of proteins, nucleic acids, lipids, and other biomolecules, in yellow or the physical properties of EVs such as density, morphology, charges or biomechanical properties, in blue. EVs are exhibiting a large variety of features in their composition and their physical properties.*

Defining the identity of EVs remains a significant challenge due to their heterogeneity and the overlapping physical and molecular characteristics with other biological nanoparticles. A proper EV characterization is mandatory to guarantee the reliability of the results in the EV research, especially considering the technical challenges of their isolation and the potential presence of contaminants[18]. The importance of this step is also underlined by the definition of specific guidelines by the International Society for Extracellular Vesicles (ISEV)[7]. There are multiple approaches to characterize parameters of an EV sample, such as concentration, size, morphology, composition, and function. These processes are usually based on different and complementary techniques, each of which is tailored to evaluate specific features. However, the heterogeneity of EVs, combined with the limitations of analytical instruments used in the EV field, makes their characterization challenging and prone to high variability[19]. Variability can arise from a lack of protocol standardization and poor interoperability between instruments. Bachurski et al.[20] demonstrated that differences in instruments' detection limits and in analysis pipelines, along with EV heterogeneity, significantly impact accuracy and precision of measurements. Arab et al.[21] showed the importance of using orthogonal methods of characterization. These findings underscore that different analytical approaches that analyze the same parameter of an EV preparation can give discordant and sometimes non-reproducible results. Furthermore, the outcomes can be also affected by lack of reproducibility and standardization of experimental protocols and instrument setting[22]. There is a pressing need for advanced characterization instruments capable of reducing variability and enhancing the reproducibility of results through multiparametric analysis.

Bulk approaches, including -omics techniques, Nanoparticle Tracking Analysis (NTA), Pulse Sensing, Light Scattering, colorimetric assays, and Raman spectroscopy can leverage analytical and biochemical properties of EVs. These characterization methods provide insights into key EV population parameters such as concentration, size, charge, molecular weight, composition, and purity. While some of these methods have been adapted to approach single-EV measurements, they often lack the sensitivity required to detect individual vesicles. On the other hand, the development of single-vesicle techniques, defined as methods enabling detailed characterization of individual vesicles at the nanometric scale, has opened new possibilities for investigating EV heterogeneity[23–25]. These advanced approaches can complement more usual methods, thus providing comprehensive information on the EV biology. They shed light on different aspects of the EV origin, their content, and their biological function, ultimately facilitating also their translation towards the clinical setting[26]. These single EV analysis methods will be presented and discussed in the following.

## I - Going down to single EV analysis
### Challenge of EV analysis
Studying EVs is challenging due to their heterogeneity in size, composition, and function. Definition of subpopulations based on a single parameter often fails to capture the overall diversity of EVs, thereby concealing valuable information and often not correlating with distinct biological functions[27]. Therefore, single EV characterization is crucial to understand their heterogeneity, to improve sensitivity and specificity of analysis, to access markers colocalization and to enable precise quantification. Among the common single-EV techniques, flow cytometry[28] is one of the most widely used, but has limitations in sensitivity, size detection limit and background noise. Other strategies, including interferometric imaging, electron microscopy, scanning probe microscopy and optical microscopy, allow the multiparametric profiling of the single vesicle, providing a comprehensive characterization of the EV samples. Integrating diverse analyses into a single technique reduces costs, minimizing experimental time and analytic variability while enhancing reproducibility and predictive values[29]. Multiparametric single EV analysis is therefore essential for robust EV characterization and support the advancement of the EV-based approaches in both the therapeutic and the diagnostic application.

### Microscopy for single-EV imaging
Among techniques for EVs analysis, single-vesicle microscopy approaches have been indicated as a fundamental step in the ISEV characterization guidelines[7]. The different imaging approaches for EVs are listed in Table 1.

**Table 1: List of commonly used imaging approaches for single EV imaging and characterization.**

| Method | Associated techniques/abbreviation | Measurable characteristic |
| --- | --- | --- |
| *Electron microscopy (EM)* | Transmission electron microscopy (TEM), Cryo-TEM, immune gold, negative contrast staining, Scanning electron microscopy (SEM) | Size, morphology and , biological markers (using gold-labelled antibodies) |
| *Atomic force microscopy (AFM)* | Contact AFM, tapping AFM, non-contact AFM, liquid or air AFM | Size, morphology and biomechanical properties |
| *Interferometric microscopy* | Single Particle Interferometric Reflectance Imaging Sensor (SP-IRIS), Interference scattering microscopy (iSCAT) | Size, morphology and , biological markers |
| *Fluorescence microscopy* | confocal laser scanning microscopy (CLSM), total internal reflection fluorescence (TIRF) | biological markers |
| *Super-resolution Microscopy* | SMLM, PALM, Direct stochastic optical reconstruction microscopy (dSTORM), stimulated emission depletion microscopy (STED), Super-Resolution Structured Illumination Microscopy (SR-SIM) | Size, morphology and biological markers |

Transmission electron microscopy (TEM), especially the negative staining approach, is one of the most used techniques to provide information about the size, shape and structure of single EVs. In electron microscopy, the light beam is replaced by electrons, decreasing the theoretical resolution limit to less than 1 nm. The contrast on the image is provided by adding a staining solution in the sample (usually Ammonium molybdate or Uranyl acetate). This approach is characterized by relatively easy and fast sample preparation. However, as the electron beam has to be under vacuum to reduce electron interaction with air, the sample must be dehydrated. This preparation procedure leads to shrinking and deformation of the EVs, with the commonly observed cup-shaped morphology being a technical artifact[30,31]. Cryo-TEM, which avoids traditional TEM sample preparation by relying on vitrification of water, has been adapted for the imaging of EVs and has gained increasing interest in the field. In fact, sample manipulation is very limited and, avoiding dehydration and fixation, it is now considered as a more optimal approach to study the morphology of the EVs. However, CryoTEM microscopy is still not easily accessible due to the need for specialized equipment and trained operators. In addition, only EVs with a size lower than 500 nm can be imaged, and those larger than 200 nm can appear flattened into ellipsoids in the vitrified film. Despite these important advantages, EM also has some general drawbacks in the EV analysis, including relatively low throughput and challenges in achieving statistical significance[32].

Standard AFM is based on a cantilever with a sharp and narrow tip used to scan the surface of the specimen. Following the encounter of the sample or the interaction between the tip and the surface, the tip is deflected, and this movement is registered by a photo diode that then allows the reconstruction of the surface of the sample. As AFM doesn't employ any beam irradiation, lenses or staining, it is not limited by diffraction or other optical and chromatic aberrations. However, it can be susceptible to a dilation due to convolution between the tip radius and the sample. In the EV field, its effectiveness has been proved in several studies as it can not only evaluate the size and the shape of EVs in their native conditions, but it can also provide information about the concentration, the biomechanics and biomolecular features of EVs[33]. However, the throughput of this approach is not usually sufficient to analyze a significant number of EVs with routine analysis[34]. In addition, the EV capture onto the surface, the sample preparation, humidity and temperature, the force applied on the sample, and the scan rate are all aspects that can affect the outcomes[35].

Interferometric microscopy has emerged as a powerful technique for single extracellular vesicle (EV) imaging. Single-particle interferometric reflectance imaging (SP-IRIS), one of the prominent methods, enables label-free detection of individual EVs by capturing them on a reflective surface and analyzing the scattered light to measure their size and binding dynamics, with high sensitivity, spatial resolution allowing sizing (~50 to 100 nm). This technic has the benefit of being label-free[36,37]. Interference scattering microscopy (iSCAT) can be adapted to EV characterization[38] for label free size measurement with high sensitivity. Interferometric plasmonic imaging, integrates interferometry with plasmonic sensing to offer excellent sensitivity for detecting small EVs. This approach further enhances the ability to perform highly sensitive, label-free imaging of EVs, allowing for more detailed analyses of their behavior in various biological contexts[39,40]. Despite their strengths, interferometric techniques have certain limitations. They are often sensitive to contamination, not well-suited for colocalization analysis of multiple biomarkers on individual EVs and it typically requires complex setups with specialized equipment and expertise[41,42].

In addition to these techniques, optical microscopy has been used for decades to study EVs and their interactions with cellular structures at the subcellular level[43]. Traditional fluorescence microscopy enables biologists to selectively label and examine cellular components with high sensitivity in fixed and living samples. The optical-based approaches, especially the fluorescence microscopy, have many advantages over the other microscopy and biochemical methods. Firstly, as it relies on a light beam composed of different wavelengths, optical microscopy enables simultaneous imaging of multiple targets (fluorophores), allowing for multiplexed investigation of the specimen. In addition, optical microscopy typically enables real-time or rapid analysis of samples. For instance, fluorescence microscopy allows the imaging of many different targets (proteins, lipids, nucleic acids, ions etc.), thanks to the possibility of using several fluorophores at once, in real time (milliseconds resolution) and in complex environment[44]. This rapid imaging of multiple molecules or biomarkers in native samples is particularly valuable from a clinical perspective[45]. However, traditional microscopy techniques lack the spatial resolution required to accurately visualize the structure, morphology, size, and surface features of individual nanometric EVs. Constrained by the diffraction limit[46], traditional microscopy methods are inadequate to visualize the size of single EV. Recent advances in super-resolution microscopy have transformed cell biology, enabling systematic nanoscale analysis of biological systems with a resolution down to the tens of nanometers. These developments open new opportunities for detailed exploration of EVs, shedding light on their heterogeneity and functional properties at an unprecedented level of precision[47,48].

Here, we reviewed the current applications of super-resolution optical imaging techniques to image and characterize the EVs, discussing some of the limitations associated with the technology as well as its future potential applications that could pave the way for new advances in this field.

## II - Super-resolution microscopy in the EV field

Developed in the early 2000s to resolve biological structures below the diffraction limit, Super-Resolution Microscopy (SRM) is fundamental to image single EVs and provide multiparametric information on the whole EV population. SRM can reach a lateral resolution of few tens of nanometers, thus allowing the visualization of nanometric biological structures[49–51]. From its beginning, numerous SRM techniques have been developed and are currently available as commercial or open tools with or without specific analysis software[52]. The resolution achieved by various SRM techniques, along with their sample type restrictions, general adaptability, and usability, can differ significantly across approaches[53]. With its exceptional resolving power, SRM holds significant promise as the optical microscopy method of choice for exploring the nanometric world of EVs, despite being at the early stages of application in the EV field. SRM has been employed for decades in cell biology and several techniques have been optimized, but in the EV field, only three super-resolution approaches have been applied so far: Single Molecule Localization Microscopy (SMLM), STimulated Emission Depletion (STED) microscopy and Super-Resolution Structured Illumination Microscopy (SR-SIM). In this review, we will describe these approaches, providing a concise discussion of each technique's fundamental principles, advantages, and limitations. Our goal is to equip EV researchers with essential knowledge to choose the most suitable method for their specific needs.

- **Single Molecule Localization Microscopy (SMLM)**

When observed under a fluorescence microscope, a single fluorophore appears as a diffraction-limited intensity distribution, dictated by the point spread function of the optical system. When captured by a detector, the pixel pattern (diffraction spot) can be deconvoluted thanks to a 2D Gaussian fitting to recover the precise localization of the emitter with a sub-diffraction resolution that is directly linked to the fitting error (Figure 2.A). The working principle of SMLM is based on the imaging over time of random subsets of fluorophores that are spatially and temporally confined: close fluorophores need to be observed successively and not simultaneously (Figure 2.E) in order to be resolved at sub-diffraction level[54] (Figure 2.C). Thus, SMLM needs fluorophores that can alternatively be in a non-emitting state, also called dark state, or in a fluorescent, emitting state[55] (Figure 2.B-C), a principle that differs slightly in the case of DNA-PAINT (Figure 2.F). The ability of a fluorophore to switch between these two states, known as photoswitching or blinking, is not a characteristic shared by all dyes used in fluorescent microscopy. Being beyond the purpose of this review, a comprehensive recap of the fluorophores and all the different techniques available for SMLM can be found in literature[56,57]. In SMLM, the final image is obtained by the acquisition of hundreds to thousands of frames and in which only few fluorophores actively emit photons[58]. The camera then captures and registers the photons emitted by each single dye molecule and the distribution of the intensity of each blinking spot is then computed and precisely localized. In the final step all the localizations are merged and combined to obtain the final super-resolved picture[59] (Figure 2.E). The SMLM is the most used super-resolution approach in the EV field and three different techniques have been developed so far: direct Stochastic Optical Reconstruction Microscopy (dSTORM), Photo-Activated Localization Microscopy (PALM) and, even if in a lesser extent, DNA Points Accumulation for Imaging in Nanoscale Topography (DNA-PAINT).

  o *STORM / dSTORM*

Stochastic optical reconstruction microscopy (STORM) can reach nanometer accuracy, but requires pairs of photoswitchable dyes, such as activator-reporter dye pairs[60]. Developed in 2008 by Heilemann et al.[61], dSTORM is a relatively easier technique thanks to the commercial availability of switchable fluorophores[62]. In this approach, labelled antibodies are used to target specific antigens. Given the potential proximity that can be reached by two adjacent antibodies, the control of the photoswitching process of the dye is fundamental in the dSTORM. This is usually assured by appropriate imaging buffers, that optimize the blinking process and ensure the activation of only a small subset of fluorophores, avoiding as much as possible their irreversible bleaching (Figure 2.B). This can be obtained thanks to the addition of thiol groups (e.g. β-mercaptoethylamine) and of enzymatic or chemical oxygen scavengers, responsible for the oxidation of the fluorophore and its bleaching[63–65]. Parallelly, the proper laser intensity for the excitation of the dye should be optimized to maximize the rate of fluorophores that get into the dark state (non-emitting) and their subsequent blinking, while minimizing the photobleaching. With an optimized dSTORM approach, usually it is possible to reach a resolution of 15-20 nm in the x-y axis[50,58]. In addition, depending on the instrument used, it is possible with dSTORM to perform multicolor and 3D imaging of the sample, even if the resolution in the Z-axis is usually lower, about 50 nm[50,63]. The main disadvantages of this approach include the high sensibility to vibration and to the quality of the labelling procedure. In fact, the label density dramatically affects the precision of localization, and a high efficiency of labelling of the targets is required[63]. In addition, the chemicals used in the STORM buffers and the oxygen scavenging generally prevent this technique from being used in live imaging[66]. Lastly, the use of antibodies, that have a typical size of about 10-14 nm[67], should be considered when quantitative measures on nanometric-sized structures, as the smallest EVs, are performed. They can potentially lead to an overestimation of the EV size due to the introduction of significant distance between the fluorophore and the actual target[68].

- o *PALM*

Differently from dSTORM, PALM is based on the detection of single genetically engineered photoactivatable fluorescent proteins expressed in the biological system. In this approach, UV laser light with appropriate wavelength and with finely tuned intensity is responsible for the conformational change and the activation of the fluorescent proteins. Once activated, the fluorophore can be excited by a second laser with a different wavelength, thus emitting photons. This on/off cycle, that can be single or repeated multiple times before photobleaching occurs in all the fluorescent molecules across the whole sample. As in dSTORM, each single emitting event is recorded on the camera and finally reconstructed into the super-resolved image by fitting the PSF of each individual event[69]. Differently from dSTORM, the PALM approach is compatible with live imaging as no harmful chemical reagents are needed, even if this is adaptable only to slow cellular processes due to the low quantum yield[62], leading to a low signal to noise ratio. In addition, by using genetically tagged proteins, the efficiency of labelling in PALM reaches 100%, overcoming a potential limitation of dSTORM approach[63]. However, these engineered proteins can also lead to some disadvantages. First of all, their function could be affected by the covalent binding of the dye to its structure[70]. Another key difference is that in PALM, each fluorophore is imaged only once, which minimizes the impact of photobleaching on sensitivity. This reduces the likelihood of spontaneous blinking from adjacent fluorophores. Since each fluorophore blinks only once, PALM is well-suited for quantifying the exact number of fluorescent molecules[71], making it a more quantitative technique compared to dSTORM. However, this is usually translated in a lower choice of photoactivable fluorophores and fewer photons emitted per molecules than in dSTORM, that can affect the accurate localization of the target[72]. Despite this, in optimal conditions PALM imaging has a similar resolution to dSTORM of about 15-20 nm in the lateral plane and 50-75 nm in the axial plane[73].

- o *DNA PAINT*

Specifically designed to overcome the limitations of the other SMLM approaches, the DNA-PAINT working principle is based on freely diffusing dyes linked to a DNA strand that transiently interact with the sample, which leads to a more straightforward experimental procedure[74]. As suggested by the name, this technique employs two short complementary DNA molecules, usually 8-10 nucleotides long, called docking and imager strand (Figure 2.F). The former is bound to the biological target through different mechanisms (DNA-conjugated antibodies or direct coupling to nucleotide sequences of interest[75], while the latter is conjugated to an organic dye, and organic dyes can freely diffuse within the buffer. While diffusing, the dye of the imager strand can emit fluorescence, but it is not localized by the system, as in the duration of a single frame (typically milliseconds) the molecule diffuses over a high number of camera pixels. On the contrary, the docking strand can transiently bind the complementary sequence, and in this case the fluorescence emitted by the static fluorophore is detected, resulting in the generation of PSFs and the precise localization of the molecule[76]. In DNA-PAINT, both the duration of the binding and its frequency can be easily controlled, allowing the user to finely tune the blinking process. If the binding depends only on the stability of the two strands DNA sequences (which can be optimized by acting on the nucleotide length and sequence, GC content or the temperature), the frequency is affected by the influx rate and the concentration of the imager sequence. This high control of the blinking kinetics ensures that the imaging does not rely on the biophysics of the fluorophore or the illumination setting[74]. This in turn makes a higher number of fluorophores adaptable for DNA-PAINT compared to the other SMLM approaches. Another important advantage of this technique consists in the possibility to perform multiplexing when orthogonal docking strands are used[76]. In this approach, many different imager strands are cyclically and sequentially introduced to the system, allowing the detection of tens of targets recreating a multiplexed image[77]. The fine tunability of DNA binding and blinking events, combined with an extremely low degree of photobleaching[78], allows an accurate quantitative analysis by molecule counting[79]. Thanks to these features, DNA-PAINT assures the imaging at an even higher lateral resolution compared to the other SMLM approaches[80], that can reach ~1 nm in optimized in vitro structures[74] and 5-10 nm for biological structures[81,82]. Despite these advantages, the imaging time of this approach is greatly higher, in the hours' time frame, compared to the others, that typically are in the order of tens of minutes maximum[74,78]. A second drawback is related to the potential background signal deriving from the free imager strands which forces the users to illuminate the samples with optical sectioning techniques, such as TIRF, HILO or light-sheet microscopy[83,84] which need to be taken in account when performing 3D reconstruction[85]. In addition, the live imaging can be very challenging, as the prolonged and numerous cycles of infusing the imager strand into living cells can be detrimental for their survival.

- **STimulated Emission Depletion (STED) microscopy**

STED microscopy was the first super-resolution microscopy technique conceived and developed in 1994[86]. The STED works in a confocal setting and its working principle is based on the minimization of the excitation area and consequently of the number of fluorophores that actively emit fluorescence. To this end, two concomitant and concentrical lasers are used, one confocal excitation laser and a second one, with a red-shifted wavelength, responsible for the quenching of the fluorophores at the periphery of the excitation area[65]. The result is a very narrow circular excitation area in which the molecules emit fluorescence[87]. The whole sample is then wholly scanned and when the rate of the STED process rate is similar or higher compared to that of the spontaneous deactivation of the excitation state of the fluorophore, the infringement of the diffraction limit is obtained (Figure 2.G). The final super-resolved image is then reconstructed by combining all the frames obtained with the scanning process. With recent STED systems a typical lateral resolution of 50-60 nm is achieved, with a slightly higher value in the z-axis (~100 nm)[50,88]. Advantageous features of the STED approach include the relatively easy to use, as it can be implanted as an add-on of confocal microscopes, the possibility of multi-colors imaging and the ability to indifferently image both recombinant proteins and tagged antibodies[89–91]. In addition, compared to SMLM techniques that require several thousands of frames, STED is characterized by a lower acquisition time and less image processing and analysis processes[87]. To obtain enough signal to reconstruct the super-resolved image, the fluorophores need to undergo numerous ON/OFF cycles in this approach, contrarily to SMLM techniques in which every single photon counts for the final picture[92]. Another potential issue related to the low number of photons is the photobleaching caused by the high power of the lasers that can prevent the necessary repeated activation cycles of the dyes[93].

- **Super-Resolution Structured Illumination Microscopy (SR-SIM)**

The SIM working principle uses the spatial frequencies contained in an image: high spatial frequencies correspond to close emitters and thus reflect small details of an image, while low spatial frequencies correspond to large structures (Figure 2.H). As in every optical setup, optical components such as lenses prevent from collecting the high frequencies of a sample. The SR-SIM approach addresses this by using a periodic interference light pattern with high spatial frequency, close to the diffraction limit, to illuminate the sample instead of a uniform field as in conventional microscopy[94]. The production of this periodic patterned illumination with the sample allows to shift the spatial content collected by the microscope. The sequential illumination of the object with several grids of different phases and orientations allows the recording of high spatial frequency. The super-resolved image can then be computationally reconstructed, with an approach based on the Fourier transform[95]. As can be easily guessed, the crucial parts in the SIM approach are the structured illumination and the computational reconstruction of the super-resolved image. Contrarily to other imaging techniques, no additional preparation or particular chemical environment are needed, which leads to several advantages. Firstly, SR-SIM can be easily applied to specimens that have been prepared for conventional fluorescence microscopy. In addition, most of the available fluorophores can be employed, the resistance to photobleaching is relatively high and multiple fluorophores can be simultaneously used[53,96]. SR-SIM is also one of the election super-resolution approaches for live imaging. However, the resolution gain is relatively lower than SMLM approach as with the most recent advancement, it can reach a lateral resolution of ~100 nm[95]. The 3D imaging is also possible, even if being based on widefield microscopy and the specific illumination pattern used, its performance is even lower (~300 nm)[97]. SR-SIM is also affected by both the characteristics of the sample and the imaging conditions, therefore a relatively long optimization work is usually needed.

The different characteristic of the SRM technics are listed in Table 2.

**Table 2: Summary of the presented SRM methods and their characteristics.**

| | | Concept | Resolution | Probes | Photo-bleaching | Acquisition time | Post acquisition processing |
|---|---|---|---|---|---|---|---|
| widefield fluorescent microscopy | | | >200nm | Fluorescent probes | | instant | No |
| SMLM | STORM/dSTORM | localization with photoswitching | ~15nm | Photoswitchable dyes and blinking fluorophores | Low | long (minutes) | PSF localisation |
| | PALM | localization with photoswitching | ~15nm | Photoactivable fluorescent proteins | Low | long (minutes) | PSF localisation |
| | DNA PAINT | localization with photoswitching | ~1nm | Fluorescent DNA probes | Low | very long (hours) | PSF localisation |
| STED | | Stimulated emission depletion | ~50nm | Fluorescent probes | High | short (seconds) | No |
| SIM | | Sinusoidal patterned illumination | ~100nm | Fluorescent probes | High | medium (seconds) | Fourier transforms |

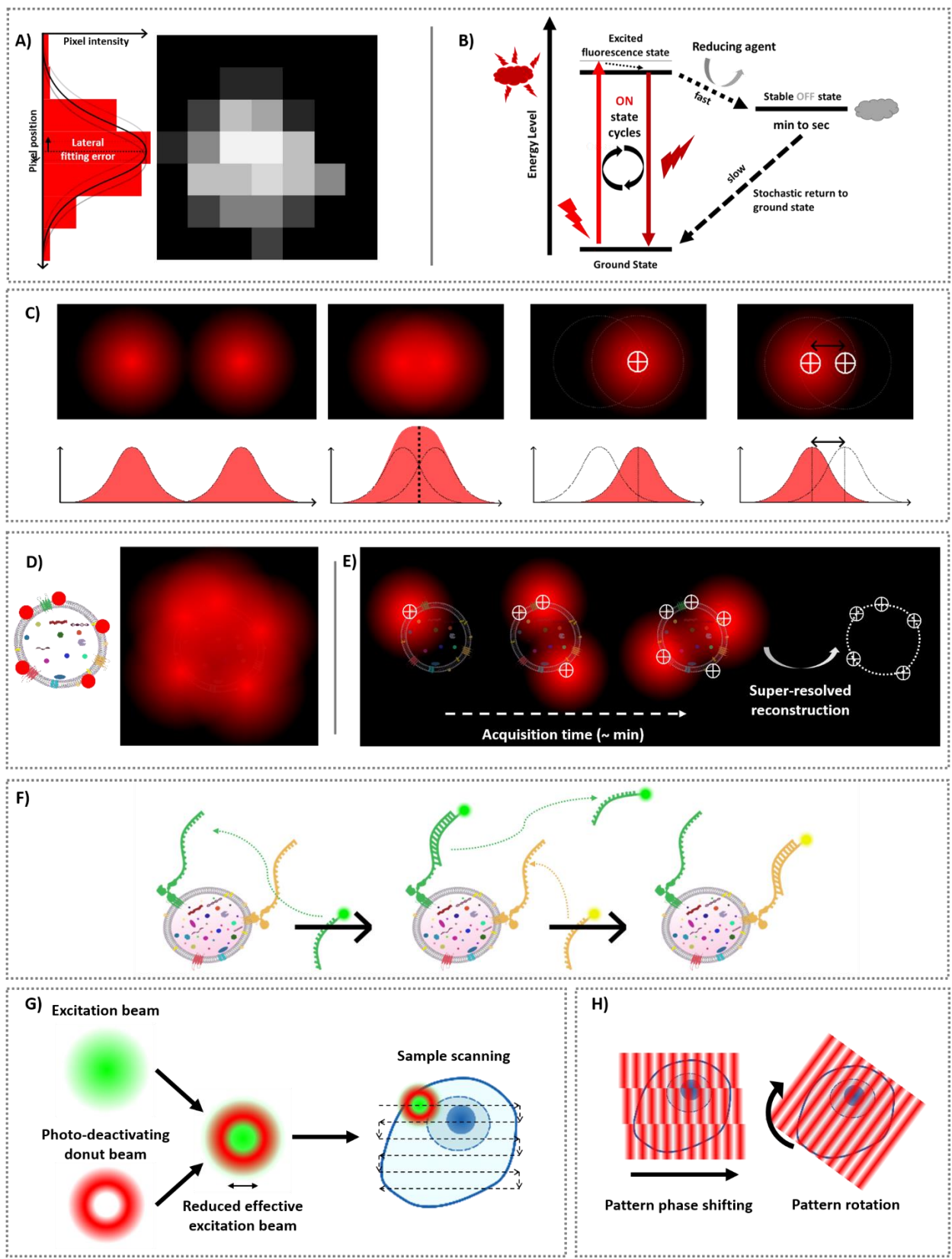

Figure 2: **EVs imaging by super resolution microscopy**. A) Visualization of a Point Spread Function (PSF) on the camera (right) and gaussian fitting of the PSF pixel intensity (left); the maximum of the gaussian gives the position of the fluorophore, the uncertainty on the fitting induces a sub-diffraction lateral localization uncertainty corresponding to the SMLM resolution limit. B) to C) Principle of dSTORM microsopy. B) In dSTORM, single emitters can be observed by making cyanine dyes blink. The Jablonski-Perrin diagram shows that the dye, in active ON state under laser illumination can be turned into a stable non-emitting OFF state thanks to a reducing agent in the buffer. The dyes can then return stochastically under long time constants to the ground state and emit light for a short time until it is brought to the OFF state again. C) Reaching a sub-diffraction resolution with SMLM: emission of two well separated fluorophores (left). If the fluorophores are closer than the diffraction limit, their PSF overlap and the two fluorophores can't be resolved (second from left). If only one of the two fluorophores is fluorescent, the center of its PSF can be localized with a sub-diffraction resolution (third from left). If the first fluorophore switches off and the second switches on, the second fluorophore can be localized as well; this way, the two fluorophores can be separated with a sub-diffraction resolution (right). D) to E) Application to an EV decorated with membrane dyes. D) Wide-Field fluorescence microscopy of an EV, its sub-diffraction size doesn't allow its characterization while all fluorophores are emitting at the same time. E) Blinking process of the fluorophores on the EV, detection and localization of isolated PSF over time which allow, after reconstruction, the observation of the EV with a sub-diffraction resolution. F) Principle of DNA-PAINT microscopy for single EVs imaging: DNA-PAINT relies on the transient binding of short fluorescently labeled DNA strands to complementary DNA targets attached to the feature of interest of the EV. G) Principle of STED microscopy: super-resolution is obtained by combining the usual excitation beam to a concentric donut beam that will photo-deactivate fluorophores at the periphery of the excitation beam, resulting in a narrower sub-diffraction excitation beam; the sample is then scanned with the effective super-resolved beam. H) Principle of SIM: the excitation beam is a periodic pattern of high spatial frequency; this patterned illumination will allow to detect smaller objects but on a partial image; full imaging is then obtained by acquiring successive images at phase shifted and rotated pattern.

### III - Key steps of super-resolution microscopy for single EV detection

The studies discussed in the previous sections highlight the significant challenges associated with imaging single EVs. When using super-resolution microscopy for EV imaging, several key obstacles arise, including optimal sample preparation, effective EV immobilization, the appropriate selection of markers, compatibility with blinking buffers, and the complexities of image analysis. To address these challenges, numerous studies have focused on developing new technical solutions aimed at simplifying and improving the protocols for EV preparation and analysis.

**Sample Preparation:** Sample preparation is a crucial step in super-resolution microscopy for imaging EVs. It must ensure minimal loss of vesicles, prevent aggregation, and maintain the structural integrity of the EVs. Preparation techniques often involve trade-offs between achieving high purity, maximizing yield, and preserving the vesicles' native structure. In the case of nanometric structures like EVs, the presence of contaminants poses a significant challenge. Contaminants could be a variety of biological entities like membrane debris, cluster of antibodies, small proteins or membrane/proteins aggregates, high-density lipoproteins (HDL), low-density lipoproteins (LDL), and very low-density lipoproteins (VLDL). These unwanted particles are often similar in size and/or density to EVs[98,99], complicating both the isolation process and the imaging workflow.

There are various preparation techniques that can help to reduce biological and imaging background noise, minimizing nonspecific binding and increasing recorded signal. For instance, the introduction of gold nanopillars can be used to capture EVs in a confined area[100], which enhances the detection of proteins of interest. This technique has been shown to significantly decrease background noise and nonspecific binding, thereby improving the overall quality of imaging. Building on expansion microscopy traditionally used for cells, an innovative approach involves embedding EVs within a swellable gel to enhance spatial resolution physically[101]. The gel's electrolytic properties cause fluorescently labeled exosomes to undergo isotropic linear expansion, increasing the spatial resolution of SRM by a factor of 4.6. This expansion allows for detailed observation of densely packed proteins on the EV membranes, providing nanoscale insights into EV structures critical for cancer diagnosis and treatment. However, this approach offers only modest improvements and remains constrained by the diffraction limit.

**EV Immobilization:** As complete imaging acquisition can usually take up to tens of minutes with SRM, effective immobilization of EVs is crucial for imaging at the nanometric scale. Considering the dimension of the EVs, any minimal movement during the acquisition can lead to an incorrect reconstruction of the EV structure. Immobilization strategies need to maintain EVs in place during the full imaging time without distorting their native morphology. If SRM has been applied in EV suspensions for high resolution localisation[102], another effective approach involves immobilizing particles on a glass slide for characterization. Immobilization provides greater resolution by reducing particle movement and allows imaging of a larger number of EVs in the same optical plane. Common approaches include either nonspecific interactions[103] or specific immobilization on surfaces coated with antibodies or ligands that target EV surface markers, as described in Figure 3.A. A commonly used immobilization method leverage electrostatic interaction through poly-L-lysine coatings[104–106]. This approach is relatively simple and can capture a broad EV population; however, it does not guarantee strong or stable immobilization, which can limit imaging quality. Alternatively, biotinylated molecules, such as BSA or polymers[104,107], can be applied to the imaging surface to achieve selective immobilization after thorough cleaning of the surface. In this strategy, EVs can be either biotinylated directly[108,109] or selectively captured by biotinylated antibodies[103,107], that bind to specific proteins or lipids on the EV surface, thus enriching for particular EV subpopulations. Other methods involve antibody coatings without biotin[107,110], further diversifying options for selective EV capture. Different approaches that have been previously developed could be adapted to SRM to bind the EVs in a non-specific way, allowing the capture of a higher number of particles present in a given preparation[102,111]. Other innovative procedures can have a good potential, like the use of lectins to non-specifically bind the glycoproteins on the EV membrane[112] or peptides sensitive to the membrane curvature[113,114]. Finding immobilization techniques that balance stability with preservation of native properties and vesicle's structure is a key challenge in obtaining high-quality images of single EVs.

**Particle markers and fluorescent tags:** The selection of particle markers is a critical aspect of EV imaging, directly influencing the resolution and specificity of the results. Traditional markers, such as antibodies against tetraspanins (CD9, CD63, and CD81)[115], or other specific proteins of EVs[116], are commonly used for labeling EVs in microscopy (Figure 3.B). If antibodies are commonly used in super-resolution microscopy to label EVs, their performance can be influenced by a variety of factors, including their affinity toward the target, their concentrations, the labeling strategy, and the sample preparation. Selecting antibodies that exhibit high affinity and specificity for the target protein on EVs ensure precise labeling and enhance imaging accuracy. Also, their concentration is important to guarantee proper imaging. For instance, Lennon et al.[110], reported an optimal concentration for Alexa Fluor 647-conjugated reporters of 13 nM (2 µg/mL) with a dSTORM setting. If a lower concentration (1 nM) led to a decrease detection of the EVs, a higher one (30 nM) increased the background noise but not the number of the EVs imaged. Considering also that a too high dye density is related to photobleaching[117], it implies that this step is crucial during the optimization of new experimental settings. In addition, different labeling strategies can be used to detect EV-associated targets with antibodies, such as direct, indirect, and sandwich labelling (Figure 3.C.). This choice depends on the experimental design and the target protein, and are important for realizing an actual single molecule localization imaging[60]. Moreover, the relatively large size of IgG antibodies, as depicted in Figure 3.C, presents a significant challenge in the context of EVs. Small EVs are almost comparable in size to an IgG complex, which can interfere with proper labeling, create steric hindrances and significantly increase their apparent size, sometimes up to double. To bypass this difficulties, alternative labeling strategies involve direct lipid staining, targeting the vesicle membrane, the insertion of fluorophores into the EV membrane during sample preparation or the use of nanobodies—smaller, single-domain antibodies[118,119] or aptamers[120–122]. This reduces steric hindrance and allows for more efficient labeling of EVs.

Although there are numerous fluorophores available for SRM imaging, it is essential to continue the research of novel molecules and the development of innovative imaging systems. For instance, silicon quantum dots showed blinking properties and they were used in combination with aptamers, that are much small than normal antibodies, to target CD63 on the surface of EVs[123]. The development of these systems, thanks to their small size of about 3-4 nm, could represent a great opportunity to improve size estimation with the SRM software. Besides, the silicon quantum dots can blink in pure water or in PBS buffer, making them more biocompatible, even if their biophysical and imaging properties were lower compared to widespread organic dyes. Yang et al.[124] explored the design of $CsPbBr_3$-based perovskite nanocrystals that are highly photostable and tunable for super-resolution bioimaging. They achieved efficient non-specific electrostatic labeling of EVs, validated by colocalization with a lipid dye marking the EV membrane. By employing $CsPbBr_3$ nanocrystals in SRM, they attained sub-10 nm localization precision, enabling the resolution of adjacent EVs separated by only 54 nm. Huang et al. recently employed Lanthanide ion-doped upconversion nanoparticles with photoswitchable properties that are suitable in a STED setting to potentially monitor the heterogeneity of the EV subpopulation during cancer progression[125]. These molecules are very bright, not prone to photobleaching nor blinking and associated with a negligible background level in the near infrared excitation. Another interesting approach used is based on the use of molecular beacons (MBs), that were employed to target miRNAs within the EVs[111]. They are hairpin-shaped short DNA sequences that form a stem-and-loop tridimensional structure that include also a fluorophore and a quencher. When the MBs bind to the complementary sequence, they stretch the hairpin DNA, the quencher moves away from the fluorophore that can emit fluorescent light[126]. This approach guarantees a high specificity and low background noise[127] and it is suitable for the localization of their targets in a complex environment, such as the interior of cells and EVs. Interestingly, the MBs can pass through the lipid membrane of the EVs and, by using different MBs with distinct fluorophores, multiplex imaging of different targets is possible[127,128].

The inherent heterogeneity of EV populations complicates marker selection, as different EV subtypes express distinct surface proteins. This variability makes it difficult to achieve uniform labeling across all vesicles. Therefore, selecting markers that not only target the appropriate structures but are also compatible with super-resolution techniques is essential. Molecular crowding and signal-to-noise ratio issues can arise from densely packed fluorophores on small EV surfaces. Striking a balance between marker specificity, fluorophore density, and signal quality is crucial for successful EV imaging.

**Imaging buffers and acquisition parameters:** Super-resolution STORM techniques often rely on the use of imaging buffers to temporally accommodate and control fluorescent dyes emission, as discussed previously. Imaging buffer composition can affect the performance of fluorophores, potentially quenching or modifying their emission patterns. Achieving compatibility between fluorescent markers or dyes and buffer which induce blinking of fluorophores, while preserving the physiological relevance of the EVs, remains a major challenge. Blinking buffers, typically oxygen-scavenging solutions, are designed to enhance imaging stability by reducing oxygen levels[129]. Adjusting these buffers, for example modifying pH to influence dye blinking or optimizing scavenger concentration, can help stabilize the off-states of fluorophores. The most popular choice, Glox buffer, combines glucose oxidase and catalase enzymes and uses glucose as a reducing agent. Alternatives include cysteamine (MEA) and β-mercaptoethanol (BME), which are also effective in promoting consistent fluorescence for SRM.

Optimizing image acquisition parameters is also crucial, as Diekmann et al.[130] demonstrated by evaluating how imaging factors affect localization precision and labeling density, both of which are essential for data quality. They investigated the influence on localization precision and labeling density of parameters such as initial photobleaching, exposure time, laser excitation intensity, and imaging speed across a variety of blinking buffers and tagging strategies. Their findings highlighted a clear balance between image quality and acquisition speed in SRM, where longer acquisition times allowed for higher resolution.

**Image Analysis:** Finally, the analysis of super-resolution images of single EV poses its own set of difficulties. The small size of EVs complicates the distinction between individual vesicles and closely associated structures or noise, even with advanced super-resolution techniques. Accurate image analysis must account for several factors, including potential drift during image acquisition, the integration of multiple blinking events, and the reduction of chromatic aberration and image registration errors, as presented in Figure 3.D. Additionally, the heterogeneity of EVs—with variations in size, shape, and marker expression—adds further complexity to image segmentation and quantification. This diversity can lead to difficulties in accurately identifying and characterizing individual vesicles. To tackle these challenges, advanced image processing algorithms, including machine learning-based approaches, are increasingly being employed[131]. Among the commonly used approaches, Gaussian fitting enables the localization of individual molecules by fitting their point spread functions to Gaussian distributions, achieving high-resolution positioning. Clustering algorithms like DBSCAN[132,133] (Density-Based Spatial Clustering of Applications with Noise) are particularly useful for identifying groups of molecules by density, distinguishing EV signal from noise, and revealing spatial patterns within the data. Tessellation-based methods, such as Voronoi tessellation[134], divide the image space into polygonal regions, allowing for a high-precision analysis of molecular distributions and identifying boundaries between clusters. These algorithms have shown promise in automating the clustering process. However, the field is still evolving, and further development of these algorithms is needed to address the complexities associated with EV imaging[135–137]. Accurate estimation of the localization precision for each method used is essential for advancing the field[138,139].

In SRM, the quality of the final picture, and of the extracted data, can be improved by the optimization of the sample preparation and the acquisition but also of the following processing and the analysis steps. In fact, considering the high number of frames and of fluorescent molecules, the post-imaging processing directly affects final results and presents significant challenges. In this perspective, a recent pipeline for analyzing single-EV colocalization introduced an open-source ImageJ plugin called EVAnalyser[140]. This tool allows for the quantification of EV numbers and colocalization, although it currently does not provide size information. ThunderSTORM plugin[141] in Fiji allows the reconstruction of SRM data, and was also applied to EVs reconstruction[106].

In summary, the optimization of the whole process, constituted by the sample preparation, acquisition and data analysis, is critical to obtain high-quality super-resolved images of EVs and quantitative data. By carefully considering antibody choice, labeling strategy, sample preparation, and validation, researchers can maximize the sensitivity, specificity, and resolution of their super-resolution microscopy experiments.

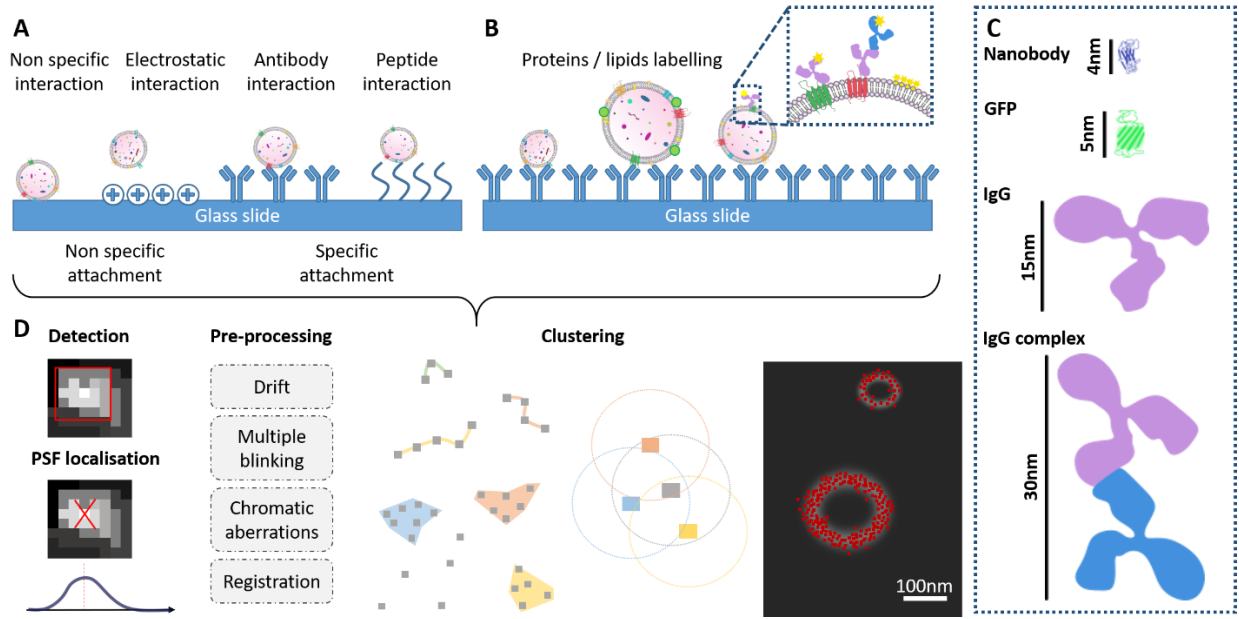

*Figure 3* **Challenges of single EV imaging by super resolution microscopy** *A) immobilization of EVs on a glass slide can be performed by nonspecific, electrostatic, antibody or peptide interactions B) Fluorophores Markers for EVs imaging targeting proteins or lipids forming an antibody sandwich around EVs C) Size scale of different fluorescent markers D) Steps of image processing from detection and point spread function (PSF) localization, recorded intensity in white and PSF localization in red, to pre-processing and clustering in order to reconstruct the EVs image with high resolution*

## IV - Imaging the morphological structure of EVs

One of the key parameters accessible through SRM is the morphology of individual EVs. SRM is among the few techniques capable of providing detailed information on the size, shape and structure of EV nanoparticles, offering unprecedented insights into their nanoscale architecture. First, we will address approaches for measuring EV size and shape, then explore three-dimensional (3D) structural characterization.

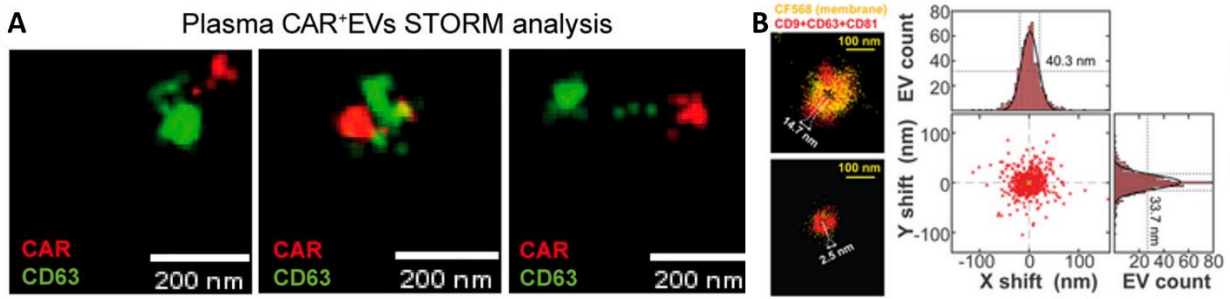

Figure 4 **Images of EVs by SRM**. A) Reconstruction of plasma EVs, captured on anti-tetraspanin antibodies coated slides, detected through dSTORM. Evs are stained with antibodies Alexa Fluor 647 - conjugated human recombinant CD19.CAR FMC63 antibody idiotype and Alexa Fluor 568–conjugated antibody CD63[142]. B) Tetraspanin-enriched EVs detected using CF568-maleimide and AF647-labelled anti-TSPAN antibodies. With the 2D distribution of relative centroid shifts for two-colour EVs (centroid of the yellow channel is assigned 0,0 position)[107]

- **EV size and shape:**

SRM allows detailed reconstruction of individual EVs to analyze both their shape and size. These analyses rely heavily on the selection of specific markers and fluorophores, which generate the signals for imaging. One strategy for achieving a more accurate size measurement of EVs is to target lipid components of the EV membrane instead of surface proteins, which are often the standard labeling approach. As lipid-targeting is generally less specific, it often results in more consistent staining, providing a uniform signal that better reflects the vesicle's true morphology. However, lipid composition is heterogeneous among EV profiles, preventing the identification of a lipid-based EV-specific marker. For instance, to quantify synaptic ectosomes, a staining approach involving wheat germ agglutinin (WGA) was preferred[110], as WGA binds glycans associated with lipids and proteins in the membrane. The result of size analysis after STORM showed a size range of 84 ± 5 nm, consistent with measurements from TEM. Another non-specific staining was also employed using Vybrant DiD, lipophilic dye with photoswitchable properties that stains the membrane of the EVs[105]. The capability of this approach in detecting and quantifying size and shape of EVs outperformed nanoparticle tracking analysis (NTA) and tunable resistive pulse sensing (TRPS). In another study[115], two membrane dyes, CM Green and CM Red, were used to stain EVs and the analysis revealed average sizes similar to those obtained by NTA. Another comparison of different methods was performed between AFM, dSTORM, MALS and microfluidic resistive pore sizing (MRPS)[104]. The size of EVs stained with Vybrant DiO, as determined by AFM and dSTORM, was consistent across both techniques. In contrast, other methods exhibited greater variability in size measurements, with significant differences observed between samples. With single-EV imaging, the authors were also able to retrieve more information about the heterogeneity and the purity of the EV samples. Various staining techniques, including WGA, Vybrant DiD, and CM Green/Red, were employed to visualize and characterize synaptic ectosomes. These approaches have enabled quantification, size determination, colocalization analysis (Figure 4.A), and even 3D reconstruction, contributing to a comprehensive understanding of these vesicles[103,115]. In addition to the lipophilic dyes, the EV size distribution could be also obtained in samples stained with conjugated antibodies directed against the three most common tetraspanins (CD9, CD63 and CD81)[143]. In Rohde et al.[144], the authors isolated two different-sized EV subpopulations through differential centrifugation and compared the size measured with SRM, electron microscopy, ExoView and NTA. The reported results showed similarity between the data obtained with SRM and those from the electron microscopy. The average EV size from NTA was however higher than with other methods and with a lower consistency, as already demonstrated. In a recent study, Saftics et al.[107] developed an innovative assay termed Single Extracellular VEsicle Nanoscopy (SEVEN) for characterization and quantification of EVs. As represented on Figure 4.B, labeling was performed by targeting not only EV tetraspanins but also specific markers originating from the producer cells[145]. SEVEN requires only a minimal sample volume (hundreds nanoliters), and provides robust and reproducible quantification comparisons among EV subtypes from plasma samples in terms of count, size, shape, and molecular composition. The EV diameters obtained through SEVEN imaging, coupled with Voronoi tessellation analysis or Extracellular Vesicle Spatial Clustering of Applications with Noise (EVSCAN algorithm, developed by Nanometrix Ltd), along with the circularity measurements, were consistent with those acquired by TEM. As expected, the sizes were slightly smaller than those determined using NTA. dSTORM was also used to assess the size of bacterial EVs[146]. Interestingly, the authors compared the size of batches of bacterial EVs stained with different approaches (Nile Red, WGA, and two bacterial proteins) reporting comparable size around 110 nm. In addition, they also measured the EV size in SEM and TEM images, which appeared slightly smaller due to absence of dyes/antibodies and the shrinkage typical of the transmission electron microscopy due to dehydration.

Clustering to reconstruct EVs with nanometric resolution provides access to their size and enables the visualization of their shape. However, correlating the measured particle size with data obtained from other instruments is challenging due to differences in the parameters being measured, such as hydrodynamic radius, gyration radius, and geometric radius[147]. Image reconstruction itself is a critical point of any measurement of the imaged particles. SRM studies commonly employ either proprietary software or custom-developed algorithms to first cluster and then measure vesicle diameters, although several commercial options are also available. The accuracy of clustering and sizing software used for super-resolution data analysis is fundamental to guarantee reliable data. However, diverse software packages might employ slightly different algorithms and parameters, leading to differences in the manipulation, interpretation and quantification of the data. To mitigate clustering artifacts, control experiments using well-characterized samples as standards[148] or negative controls should be always included. This underscores the importance of detailed reporting on data analysis software, including algorithms and parameter settings.

Last but not least, experimental parameters can also affect the size determination, as previously discussed. For instance, even if the lipophilic dyes can distribute almost evenly within the surface lipids, a too high density of the dye can result in the over-clustering, thus leading to inaccuracies in the localization of individual molecules[130] and overestimating the size distribution. When antibodies are used, the staining may be uneven, even when targeting highly expressed EVs proteins like tetraspanins, making the clustering process more laborious. This could be even more relevant for larger EVs which, due to their higher surface area and complexity, could have a more uneven distribution of lipids and proteins across their membranes. More importantly, it has to be considered that the antibodies have a size, around 14 nm[149] that can be half of the radius of the smallest vesicles[3] and could thus not be negligible. The demonstration of this effect was experimentally demonstrated[115]. In this work, the authors were able to visualize the progressive increasing distance between a first signal from a lipophilic dye and a second fluorophore present respectively within the membrane lipids, on an engineered surface protein or on a conjugated antibody directed against a surface protein. For this reason, innovative strategies based for instance on nanobodies[150] or aptamers[100] could greatly improve the EV size determination through SRM.

A recent publication from Jung et al.[151] highlights the challenges in accurately measuring the sizes of EVs smaller than 100 nm when using 2D projections from 3D imaging data in SRM. The primary issue is that localization errors—caused by distortions in the point spread function (PSF) and the limited number of photons available—can result in inaccurate size estimations when working with such small particles. To address this, the researchers introduced a correction factor that aligns the apparent size in SRM measurements more closely with the physical particle size measured by TEM, a technique that provides size measurements without these particular limitations. This correction approach raises important questions about the accuracy of imaging techniques for nanoscale EVs and suggests that for precise measurements, especially in the sub-100 nm range, imaging must account for these localization errors to avoid misleading interpretations of EV sizes. The study implies that future advancements in EV measurement will require improved 3D localization approaches to ensure that SRM can provide reliable and consistent data.

- **3D reconstruction**

Accessing 3D information by SRM on EVs is a key advancement in the field, providing valuable insights into their structure, function, morphology and improving size determination, though it remains a challenging aspect of their characterization. In most of the reviewed works, the pictures and the quantification were performed on 2D projections of the 3D vesicles. By avoiding the dehydration steps that causes potential alteration of the EV 3D structure in many EM protocols, SRM can allow a reliable visualization of 3D EVs structure with nanometer resolution. However, like in conventional microscopy, the optical resolution is lower in the Z-axis. In the EV field, the 3D imaging has not been commonly used, thus it still needs to be optimized. In this regard, McNamara et al., reported important progresses in the 3D imaging of the EVs[103,115]. In their work, they used a cylindrical lens to induce astigmatic aberrations in the detection light path. In this way, the axial information of the fluorophores was encoded in the shape of the deformed PSF. The dSTORM pictures of EVs stained with lipidic dyes showed an elliptic shape linked to the lower resolution in Z. Interestingly, this was computationally corrected with an algorithm that considered the minor diameter of the ellipsoid correspondent to that of a sphere. Therefore, the coordinates of the non-equatorial localizations were corrected, and the resulting EV picture looked more like a sphere. Importantly, with this transformation the relative angles between the data were preserved. In fact, the protein clusters of EV markers on the surface can still be effectively identified. In a related study, Puthukodan et al.[152] combined 2D SRM with 3D tracking to investigate the dynamics of EVs entering cells. A cylindrical lens was integrated in the optical pathway, creating an elongation of the 2D signal of the particles, thereby enabling depth information. The authors then used a custom software, 3D STORM Tools, for EV localization and combined this with a modified Trackpy software[153] for 3D EV tracking. This approach allowed them to observe the spatio-temporal dynamics of EV uptake by cells. The team used dual-color SRM to achieve colocalization, visualizing both the EVs and the cell membrane.

The 3D imaging with SRM has been extensively applied to fixed cellular and intracellular biological structures that are much larger than EVs[154–156], while the 3D visualization of single vesicle is rare in literature. One explanation lies in all the difficulties encountered when it comes to imaging a single EV with this approach. All the obstacles listed so far can potentially affect the 3D reconstruction of the EVs. For instance, multiple antibodies can bind to a single vesicle, creating a sort of corona. In addition, they could act as a spacer between the fluorophore and the actual EV surface, which is even more relevant for the smaller EVs. These can lead to misinterpretation of the EV morphology. On the other hand, an antibody could also hinder the binding site for a second one if they are very close, creating steric hindrance. In this regard, the orthogonal approach used in the work of McNamara et al.[115], based on the combination of direct labelling and antibodies detection, could represent a good expedient to avoid this issue. Other technical considerations concern the choice of the best fluorophores. Molecules with high brightness, signal-to-noise ratio and quantum yield should be used, especially in those approaches such as astigmatism in which the emitted photons contribute not only to image reconstruction but also to the localization precision. Linked to the use of lenses, the apochromatic aberrations should be considered and reduced as much as possible to improve 3D imaging. The reconstruction algorithms should also be specifically optimized for this approach. They must allow effective and reliable axial compensation, maintaining the actual reciprocity of the different molecules, as shown in the discussed work[115]. This algorithm should account for the potential flattening of lipid spherical particle upon binding to the imaging surface. This process can depend on several factors, modifiable ones like the density of the capture groups, and intrinsic ones such as the particle's biomechanical properties. To conclude, 3D imaging of the EVs through SRM is still challenging but could have crucial implications in the visualization and in the analysis of the EVs. In this frame, it is important for the scientific community to start this improvement and optimization process from the beginning of its application in the EV field, so as to draw a line to be followed for future studies.

## V - SRM to access EV biological composition

Visualization of the components of EVs at a resolution of tens of nanometers is a challenge that SRM can address, enabling a deeper understanding of their identity. This high-resolution imaging provides crucial insights into the biological composition of EVs, in particular to their surface markers, cargo, and the protein corona.

- **EV surface markers**

ISEV has identified the evaluation of protein markers on the surface of EVs as a crucial step in the EV characterization process to confirm vesicle identity and detect potential contamination[157]. Moreover, the presence of proteins can shed light on cellular EV-related processes and the pathophysiological state of the secreting cell. SRM, especially dSTORM, represents a powerful and tailored tool for studying EV surface markers and their complexity. Unsurprisingly, the majority of the studies included in this review employed SRM approaches to precisely investigate membrane proteins. While some studies have used SRM as a complementary characterization tool to verify the presence of classic EV markers, others have leveraged it to link specific markers to biological and pathological processes.

- Established EV-markers: Tetraspanins

The tetraspanin proteins CD9, CD63, and CD81 are by far the most commonly used markers to confirm the presence of EVs in studied samples. Consequently, several studies relied on SRM, and in particular dSTORM, to characterize EVs by specifically detecting one of these proteins[143,158–161]. A related approach involves using a combination of antibodies targeting the three tetraspanins, where each can be labeled with a distinct fluorophore, to enrich the information obtained from imaging[140,162]. Simultaneous detection of these antibodies enables the relative quantification of these proteins across different samples or EV subpopulations[115,143,158,159,161]. This can be crucial to study EV biology and their effect on recipient cells, as demonstrated by the relevance arose around the biological role of different EV subpopulations[163]. Coupling the capacity for relative quantification in dSTORM[164] with the nanometric resolution inherent to SRM give access to data that represent a significant advancement in the field. For instance, McNamara and colleagues[115] showed that CD9 and CD81 are not evenly distributed within the lipid membrane but are part of micro-domains on the surface of individual EVs. In this study, the authors stained the samples with a lipophilic dye (CM Red) and then added the anti-tetraspanin antibodies, looking for colocalizations. Thanks to multi-channel imaging, they were able to quantify the percentage of single and double positive EVs and to visualize micro-domains. However, the EV markers are not limited to the aforementioned ones. SRM enables the localization of a wide range of markers, both physiological and pathological, found on the surface or encapsulated within EVs.

- Bioactive EV markers

The EV surface markers are not only important for the characterization of a given sample, but they can also provide information about the origin and the physiologic function of the EVs. For example, the vascular endothelial growth factor (VEGF) can be packaged and released into EVs by several cell types, including cancer cells, endothelial cells, and mesenchymal stem cells. These VEGF-containing EVs, in which the signal colocalized with CD63, can then be taken up by other cells, thus stimulating the angiogenesis process[165]. Among the other EV specific markers frequently assessed, the cytosolic protein TSG101, component of the ESCRT complex, is important for the EV biogenesis[166]. Using STED microscopy this marker has been visualized, together with CD63, in the EVs[167]. Surface markers on EVs reflect their cellular origin. For instance, the co-localization of CD9 and CD42a has been identified as specific to platelet-derived EVs, where these markers are more abundant compared to EVs from B cells, which are instead enriched in CD19[168]. In another example, antigen (HLA-G) presenting EVs colocalized with CD63 have been implicated in immune tolerance during pregnancy, as they may be involved in protecting the fœtus from the maternal immune system[161]. Another human leukocyte antigens (HLA-DR) has been found on the surface of EVs released by macrophages that are involved in presenting antigens to T cells, initiating the immune response[162]. By activating signaling pathways, EVs from T cells can enhance antigen-specific responses in target cells. Using three-color dSTORM, protein microdomains responsible for this machinery were localized in EVs with high resolution[169]. Finally, EV subpopulations with different origin were distinguish using specific cellular neuronal markers[170].

- Pathology-derived EV markers

The imaging of EV surface markers with SRM can also have diagnostic and prognostic potential as it allows the identification of specific EV populations secreted by cells in pathological conditions. SRM also enables the monitoring of changes in their composition or in their relative abundance in response to disease or infection. For instance, Lennon et al. [110] identified through SRM an enriched population of EVs expressing CA19-9 and the epidermal growth factor receptor (EGFR) in pancreatic ductal adenocarcinoma (PDAC) patients compared to the EVs obtained from healthy individuals. In another approach, the SRM was used to confirm the presence of the spike protein S2 of the SARS-CoV-2 on the EVs secreted by cells transfected with the corresponding gene[171]. In addition, the authors visualized colocalization with EV markers and the binding to ACE2 receptor. EVs secreted by HBV-infected cells can also serve as scavenger of HBV virus, protecting them from antibody neutralization because of the lack of virus antigens on their surface, as shown by STED microscopy[172]. In addition, Chen and colleagues[102] imaged and tracked cancer-derived EVs thanks to the labeling of CD63 and Human Epidermal

Growth Factor Receptor-2 (HER2) on the their surface. They were able to observe the interactions of EVs with the cell membrane. They then demonstrated that it is possible to detect several EV-associated markers in the same imaging session by using DNA-PAINT[173]. In their interesting work, the authors developed a quantitative multiplex imaging platform with which, in association with a machine learning algorithm, they were able to identify up to 4 tumoral markers (HER2, GPC-1, EGFR and EpCAM) and 2 tetraspanins (CD63, and CD81) on serum derived EVs with an accuracy of 100%. An interesting complementary approach was implemented by Yin et al.[174], who relied on quantitative PAINT introduced by the team of Jungmann[79,80] to explore the potential of EVs as biomarkers. They developed a microfluidic chip allowing the coupling of cell culture, EVs immobilization and super resolution characterization. They then used fluorescent DNA probes for staining of PD-L1 proteins on the surface of EVs, with a possible quantification of the target at the nanoscale. In addition to human-derived samples, SRM have been employed to detect bacteria-derived EVs, the so-called outer-membrane vesicles (OMVs) in biological fluids. For instance, the OMVs secreted by *E. coli* were detected using SIM and AFM methods. This imaging approach can be applied also for the visualization of mammalians, plants, and samples from other organisms[175]. Jeong and colleagues demonstrated through dSTORM that enterotoxin B, a virulence factor, is transferred within the EVs from the gram-positive bacteria cell wall before being secreted[176]. All these examples demonstrated how SRM can give fundamental insights about the heterogeneity, the phenotype, the effect and the biology of EVs just by looking at classic surface markers.

The detection of surface markers on EVs is probably the most straightforward application of SRM approaches. Despite the relatively high number of studies, there are still some important considerations to be made for further optimization and advancement of these approaches. Key areas for improvement include detection specificity, sensitivity and detection limits, reduction of background noise, and the use of appropriate control samples. The recently published work of Shihan Xu et al.[177] opens this discussion by comparing SRM imaging characterization with EV flow cytometry. They address challenges in both methods, particularly the identification of contaminants such as antibody aggregates, and emphasize the importance of rigorous sample preparation and signal processing. Their findings highlight that microscopy imaging offers higher sensitivity and resolution for single-molecule detection than flow cytometry. Most of the previously discussed studies, and in particular those that used SRM as an EV characterization tool, captured the EVs with one or more antibodies directed against the three tetraspanins CD9, CD63 and CD81. A potential bias is introduced in this experimental design as only the EVs positive for one or more of the tetraspanin markers will be captured and imaged. On one hand, this could help to address some biological questions, as it has been demonstrated that the tetraspanin profile changes according to different EV subpopulations, sampling site and biogenesis pathway[14,178]. On the other hand, it implies that EVs negative for CD9, CD63 and CD81 will not be analyzed. Studies have shown that the number of EVs captured using non-specific methods is greater than that captured with anti-tetraspanin antibodies[109]. This highlights the ongoing need for non-specific, universal markers that can exclusively and comprehensively select all EVs[7]. Moreover, the double step of immunoaffinity attachment (Figure 3B), for both capture and detection, provides a great specificity for the effective detection of the target EVs. EVs from complex samples such as biofluids can be directly imaged, without any isolation step. This can be highly beneficial for the development of innovative diagnostic tools based on this approach. Using antibodies with different targets for EV capture and detection facilitates colocalization studies by imaging EVs positive for both targets with a single fluorophore. This approach can enable multiplexing of the targets[179], while reducing steric hinderance and chromatic aberrations. It also addresses a key limitation of dSTORM, that is the parallel visualization of multiple targets. As discussed in a previous paragraph, the DNA-PAINT technique has the potential to fulfil this important task, allowing the detection of multiple proteins on the EV surface[173,180]. This study has opened the doors to the multiparametric analysis of the surface markers on the EVs through SRM, thus maintaining the benefits of the single molecule resolution and representing a great opportunity to deeply investigate the EV biology.

- **EV cargo:**

SRM has been used to image intracellular biological structures at the cellular level. Therefore, it is not surprising to find studies in literature that detected specific biological content inside single EV with SRM, even if it is not as straightforward as detecting surface markers. What is more surprising, considering usual SRM detection strategies, is that many of these studies have prioritized the detection of nucleic acids over proteins. Both nucleic acids and proteins are crucial in the EV characterization process.

The accessibility of the targets inside the vesicles is a key point of cargo analysis. In a study from Valcz et al.[181], STED microscopy was employed to target Alix and Rab7 within EVs fixed with 4% paraformaldehyde (PFA) without permeabilization. Silva and colleagues[182] engineered cell lines to load GFP into EVs by overexpressing various EV sorting proteins, without fixation or permeabilization. They quantified GFP copies in individual EVs and analyzed the distribution of GFP across EV subpopulations based on the overexpressed sorting protein. In the work of Singh et al.[116], they use a mild-detergent-based permeabilization, relying on low concentration of Triton X-100, to access the cargo of EVs and detect by SRM viral proteins of SARS-CoV-2 virus inside EVs of covid-19 patients. This approach allowed for the quantification of viral protein levels both inside and on the surface of the EVs. Surprisingly, the quantities detected inside the vesicles were comparable to those on the surface.

As previously mentioned, many papers focused on the detection of nucleic acids. One of them further supported the hypothesis of the presence of DNA both within and on the surface of EVs[183]. The authors employed native EVs that were treated or not with DNase in combination or not with a permeabilization step after fixation (4% paraformaldehyde). The EVs were identified by targeting CD63 and/or CD81, while the DNA with a nucleic acid dye, demonstrating that most of the EVs carry DNA, especially on the surface. Similar results were provided by Chetty et al. [106], focusing on CD63-positive EVs. In another study, colocalizations between a DNA dye (Picogreen) and CD63 and EFGR EV markers were found using a SIM approach after EV permeabilization[184]. In addition, the authors reported that a large proportion of DNA-positive particles were also positive for a cytoplasmic (CellTracer Far Red) and a lipophilic (DiD) dye, thus confirming the association between DNA and cell-derived EVs.

Besides DNA, other nucleic acids' species enclosed within EVs have been detected with SRM. In one of these approaches, the EVs were added with exogeneous RNA through electroporation and their protection role from RNase digestion was demonstrated[185]. Even in this case, neither fixation nor permeabilization of the EVs were reported. In other studies, the authors were able to target even the miRNAs with SRM. Chen et al. [111] employed molecular beacons that became fluorescent after the hybridization to complementary target sequence (miR-21 and miR-31). Colocalizations of the signal from EVs, stained with a membrane dye and the two miRNA targets together were detected. The cells were not fixed or permeabilized. In another work, miR-21 was enriched in EVs by transient permeabilization and then detected using a DNAzyme probe, without any fixation and permeabilization steps[186]. The authors also developed a stoichiometric assay, by which they could quantify the copy number of the miR present in each EV. Interestingly, EVs secreted by cancer cells had more copies of miR-21 compared to the healthy one. These results that were validated by RT-qPCR. This assay also showed differences in the plasma derived EVs between patients before and after chemotherapy treatment. Lastly, miR-31 was detected within GFP-positive EVs using a complementary fluorescent probe[187]. Notably, the authors did not report any fixation or permeabilization of the sample in this case either.

The accessibility for visualization of EV internal targets represents a notable technical challenge due to the small size and the intrinsically heterogeneous nature of the EVs. Fixation and permeabilization are commonly used to allow the entry of antibodies or other fluorescent probes. The addition of detergent to cell membranes can create pores with a typical size of more than 200 nm[188,189]. When applied to EVs, the detergent significantly destabilizes their membrane. Achieving a careful balance between fixation and permeabilization is therefore essential for optimal sample imaging. However, permeabilization remains crucial to ensure antibodies access to the vesicles interior. Because of their size (14 nm), only a relatively small number of antibody molecules can be accommodated within a single EV. It can affect and underestimate the quantification of the target molecules present within the EVs. Given the relevance of these steps, it is necessary to clearly state if and how samples were fixed and permeabilized to allow not only the repeatability of the experiments but also the comparison of the results between different studies. Another crucial aspect to be taken into consideration is the inclusion of the necessary controls to avoid the detection of non-specific fluorophore aggregates stuck within the EVs.

- **Protein corona:**

In biological environments such as biofluids, synthetic or natural nanoparticles interact with soluble proteins that adsorb to their surface, forming a layer. This layer, called protein corona, can alter the interactions of nanoparticles, including EVs, with other proteins or with different biological entities, such as cells[190,191]. Given the relevance of the protein corona in defining the behavior of the EVs in biofluids, it is crucial to study this process to better understand the *in vivo* biological effect of the EVs on cells. To assess the lifespan of EVs coated with a protein layer, Liang and colleagues[192] used an AlexaFluor 488-conjugated albumin to decorate the surface of EVs that were engineered with a surface marker that express albumin binding domains. With dSTORM, they detected single EVs bound to albumin molecules, reporting a robust binding capacity both in vitro and after in vivo injection. In two other studies, the spontaneous formation of a protein corona around native EVs was observed. In the first, the EVs with anti-tetraspanin fluorescent antibodies were added to a solution of fluorescent (AlexaFluor 647) albumin and imaged with dSTORM[193]. In the second one, colocalizations between vesicle tetraspanins and albumin or VEGF was reported, thus demonstrating the spontaneous formation of a protein corona around native EVs and unravelling with SRM its potential biological role[159].

Recently, there has been a growing and debated significance attributed to the protein corona within the EV field[194]. SRM has proven effective for studying protein corona, which forms spontaneously and may act as a carrier for growth factors. It can be used to detect specific targets within the EV protein corona. Multi-color SRM approaches can theoretically be applied to explore colocalization between vesicular markers (proteins or lipids) and the target of interest. In addition, using a DNA-PAINT approach could significantly expand the range of potential targets, even if the validation could be challenging. Contrarily to conventional approaches, and more easily than immuno-EM, the strongest potential of SRM lies in the effective locating of specific markers. Thanks to the nanometric resolution and by including appropriate controls, SRM can elucidate if the specific target is within the EV corona, on the lipid surface or within the EVs or even if it should be considered as a contaminant.

## VI - SRM in EV biological applications

Overall, advances in SRM in the EV field have opened up new avenues for research and promise to deepen our understanding of the role of EVs in health and disease. SRM has already been applied to various biological contexts, covering the entire EV life cycle, from biogenesis to endocytosis (Figure 5).

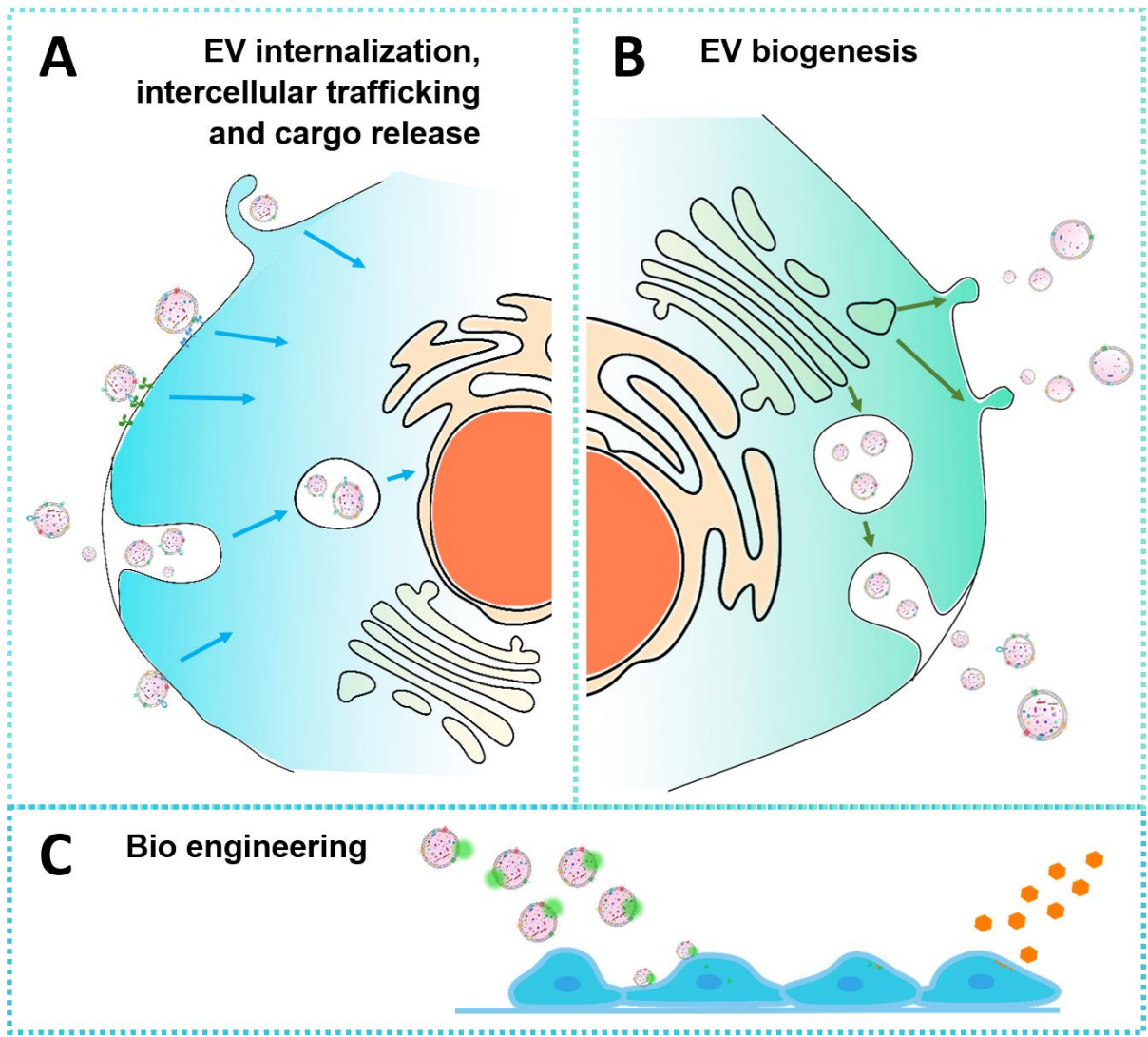

*Figure 5 **Internalization, biogenesis and engineering of EVs**. A) EVs are internalized when they reach their target cells by endocytosis, phagocytosis or interaction with the membrane, they are transported inside the cells and can release their cargo. B) EVs are released by their parent cell through the endosomal pathway or by bubbling of the plasma membrane. C) EVs can be engineered for transport of fluorescent or therapeutic molecules*

- **EV internalization, intercellular trafficking and cargo release**

The nanometric resolution achieved by SRM allows for the precise localization of fluorescent targets, enabling the study of cellular entry from an entirely new perspective. As illustrated in Figure 5.A, it includes critical phenomena such as EV trafficking, membrane fusion, cellular uptake and the fate of internalized EVs. Internalization of EVs was explored by Isogai et al.[195], demonstrating that integrins present on EV surfaces facilitate their docking onto living cells by binding to laminin. A recent work from Puthukodan et al.[152] explored the mechanisms of EV entry into cells using advanced SRM. Through 3D colocalization analysis of two-color dSTORM images, the researchers demonstrated that 25% of internalized EVs colocalized with transferrin. This colocalization is associated with early recycling endosomes and suggests that clathrin-mediated endocytosis plays a significant role in EV uptake. Building on this, Schürz and colleagues[140] quantified the EV internalization process by correlating total intracellular fluorescence with EV concentration and identifying single cells involved in the uptake. Notably, their work provided valuable insights into the kinetics of the internalization process, advancing our understanding of EV-cell interactions. Colocalization between EV and lysosomal markers was also reported, demonstrating that a large portion of the EVs was found within the lysosomes once they entered the recipient cells. SR-SIM was used to confirm the active EV internalization process by mesenchymal stromal cells[196], but also to evaluate their intracellular fate.

Interactions between EVs and various cell types were imaged using SRM to investigate cellular uptake mechanisms. de Couto et al.[197] employed dSTORM to successfully demonstrate EV uptake by macrophages via time-lapse imaging. Notably, this study did not rely on modified imaging buffers, demonstrating that dSTORM can effectively capture relatively fast biological processes, such as EV membrane uptake. In another approach, cancer EVs were actively internalized and tracked within the recipient cell[102]. The delivery of transmembrane proteins by EVs to the spermatozoa membrane was studied using membrane dye staining and targeting fusion proteins, revealing sequential interactions with different spermatozoa zones[198]. A study showed that EVs from a skin commensal fungus, linked to certain disorders, are internalized by keratinocytes and monocytes, accumulating in the peri-nuclear region to likely deliver their cargo[199].

Regarding EV fate after uptake, Chen et al. [111] used dSTORM to track the intercellular transport of EVs and their miRNA cargo after cellular entry. To address the cytotoxicity of standard dSTORM imaging buffers, they innovatively modified the buffer by substituting culture medium and reducing the levels of β-mercaptoethanol and oxygen scavengers. This enabled the reconstruction of a time-lapse video with nanometric spatial resolution and a temporal resolution of four seconds, capturing both EV movement and miRNA release into the cellular environment. Other approaches combined dSTORM and PALM to detect antibodies directed against specific markers and engineered proteins. For instance, Polanco et al. [200] reveal an unusual EV trafficking pattern, where not all internalized EVs are degraded. Instead, EVs can hijack the endosomal pathway and exploit the cell's secretory machinery to creates a persistent subpopulation with extended action and potential pathogenicity. All these findings highlight the potential of SRM to explore dynamic EV behaviors in live-cell contexts.

- **EV biogenesis**

The biogenesis of EVs, represented in Figure 5.B, is a fascinating biological process involving multiple mechanisms. Once again, SRM can play its part in elucidating these processes, with time lapse imaging, now feasible with many SRM approaches. In an advanced study, Saliba et al.[169] used planar-supported lipid bilayers with fluorescently labeled triggers and STORM imaging to visualize vesicle secretion from killer cells and quantify synaptic secretion. Complementing this, Ambrose et al.[201] provided additional insights into the mechanisms of synaptic extracellular vesicle formation and cargo selection, further elucidating the intricate processes involved in immune cell synaptic communication. SRM has also been used to study the biogenesis and the secretion of gram-positive bacteria EVs[146]. By staining various targets (membranes, cell walls, EV surface proteins), the authors linked the subcellular origin to differences in production rate and cargo composition. They identified two biogenesis types: EV secretion occurred either through local loosening of the peptidoglycan layer (lipid membrane bubbling) or explosive cell wall lysis after encapsulation and pressure increase. All these observations were possible only with dSTORM, as the presence of the cell wall prevented most of these processes from being visualized using electron microscopy.

- **Bio engineering**

As previously discussed regarding EV cargo imaging, SRM holds the unique ability to confirm the presence of specific targets on or within individual EVs. This capability makes it invaluable for quality control in the engineering and loading of EVs (Figure 5.C). Biagini and colleagues[202] employed SRM to evaluate the presence of a transgene within the EVs isolated from genetically modified zebrafish to express the oncogene RAS fused with GFP. The SRM detected GFP signal within the EVs, thus demonstrating the effective presence of the oncogene in the secreted EVs. In two different works, the enrichment of CD63-GFP on the EVs secreted by engineered cells was demonstrated through dSTORM[140,203]. As the intensity of the fluorescent signal in the EVs increased with the dose of the plasmid given to the cells, it was also possible to assess the effectiveness of the transfection. In relation to these studies, the incorporation of siRNA in lipid-based nanoparticles or the presence of fluorescent proteins on silica-based nanoparticles was demonstrated through SRM[140]. The possibility to not only verify the actual engineering but also to obtain quantitative data at single EV scale brings the SRM among the techniques that can be used to characterize EV engineering of clinical interest. The production of engineered EVs batches with relatively low variability could facilitate the development of assays by which the effectiveness and the efficiency of the process can be quantified. Allowing the multi-parametric characterization of an EV population at single particle level, SRM can decrease the variability usually associated with the characterization process that is due to the use of different instruments in parallel. Therefore, the scalability of this approach will be important in the development and validation of industrial components of not only EVs but also of other lipid-based carriers.

## VII –Perspectives on super-resolution imaging in the EV field

SRM plays a critical role in the analysis of EVs due to its ability to overcome the diffraction limit of light, which typically restricts conventional optical microscopy to resolving objects smaller than 200 nm. Since small EVs are in the range of 30–200 nm, conventional microscopy cannot resolve their structure or morphology with high enough precision, making SRM techniques essential for detailed EV characterization[107,115]. SRM, including STORM, PALM or DNA-PAINT, provides the spatial resolution needed to observe individual vesicles, their surface features, and their interactions with other molecules at the nanoscale[60,204]. This is especially crucial for understanding the heterogeneity of EV populations, as EVs vary in size, biological content, and functional properties, which traditional methods struggle to differentiate[205].

We identified three key areas for improving the use of super-resolution microscopy in EV research:

1. **Sample preparation and labelling**

The first step of sample preparation and labeling is critical for achieving successful SRM imaging. STORM relies on accumulating a large number of localizations over time to achieve high localization accuracy and spatial resolution. Consideration on the experimental parameters such as the immobilization of the object over the full imaging process, the dye used, its blinking properties, or the number of necessary localizations for reconstruction of features are crucial in STORM where precise switching of fluorophores is crucial[206]. Selection of reagents optimized for SRM is a key step, as using chemicals or dyes not tailored to these methods can lead to suboptimal results. Therefore, a greater awareness of the specific requirements and mechanisms inherent to these super-resolution techniques is essential for researchers aiming to achieve accurate and reproducible results. Control over staining protocols is also crucial, as the choice of antibodies can affect both the size measurements and the accuracy of internal staining. Researchers should consider the specific aims of their experiments and explore alternative approaches, such as DNA-PAINT, which may offer more precise or relevant results for EV analysis in certain contexts.

2. **Image reconstruction and data analysis**

Enhancing data analysis workflows is critical for improving the accuracy and reliability of super-resolution imaging. In broader biological research, SRM has progressed by refining reconstruction algorithms tailored to specific biological contexts. However, the EV field presents unique challenges: the undefined and nanometric-scale shapes of EVs are often comparable to the resolution limits of SRM. The lack of a definitive reference shape necessitates rethinking reconstruction algorithms and establishing stringent controls to validate findings. Incorporating robust quality control measures, such as tilt, focus, and PSF width analysis, ensures proper imaging setup and minimizes experimental distortions. These tools are instrumental in maintaining consistency and optimizing imaging conditions. Accurately evaluating the resolution and precision of images is crucial, not only for a comprehensive understanding of EV structures but also for studying EVs within biological processes to elucidate their functions. This ensures the reliability of observed interactions and aids in interpreting their roles in complex biological systems.

3. **Implementation of SRM in live imaging in biological relevant preparation**

SRM, with its nanometric resolution, offers groundbreaking advancements in live imaging of EVs. Unlike conventional fluorescence microscopy, SRM can precisely localize single EVs within specific subcellular compartments. However, SRM's reliance on computational reconstruction imposes limitations, such as the inability to image processes occurring faster than the acquisition time required for image reconstruction. The choice of imaging approach and buffer composition is critical, particularly for dSTORM, which traditionally employs cytotoxic buffers[197]. While suitable for rapid dynamics, these buffers limit long-term live imaging applications. Advances in cytocompatible buffers have extended recording times to several minutes[111], emphasizing the importance of balancing biocompatibility with optimal fluorophore photoswitching and photobleaching prevention. While techniques like PALM and STED are often restricted to very fast processes, SR-SIM stands out as a viable option for longer live imaging sessions, albeit with slightly reduced resolution. Multiplex imaging, crucial for investigating EV interactions with multiple cell organelles, remains challenging due to current technological limitations. Most studies employ only two fluorophores, requiring multiple imaging sessions for broader targets. DNA-PAINT shows promise in overcoming these barriers[207] offering potential for multiplexing several markers and providing comprehensive visualization of cellular structures and EV interactions.

**Toward standardization**

The field of SRM urgently requires greater transparency in protocols and data analysis workflows to advance the standardization and reproducibility of EV characterization. Clear guidelines on imaging parameters, staining strategies, and reconstruction algorithms are essential for identifying the key factors that influence EV analysis, such as size, composition, and spatial distribution. Enhanced protocol sharing would help researchers address the challenges of EV heterogeneity and ensure consistent, high-quality results across studies. In this spirit, Figure 6 introduces a template that compiles the critical parameters influencing SRM analysis. These parameters must be clearly reported in each study to allow for meaningful comparisons between studies and results. We hope this will help reduce variability and improve the reproducibility of SRM methods. This effort represents an important first step toward the standardization of SRM analysis in EV research, essential for tackling the complexity of the field.

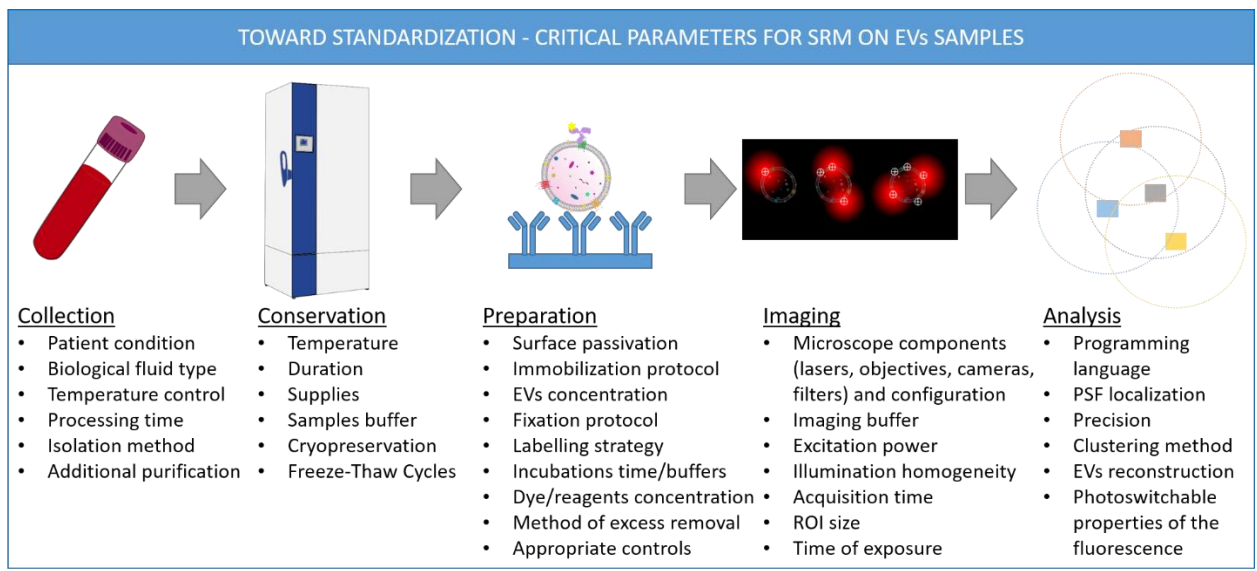

*Figure 6 **Overview of the critical parameters for SRM on EVs samples**. The SRM workflow is divided into five key steps: sample collection, conservation, preparation, imaging, and analysis. For each step, crucial points are identified that must be reported to reduce variability, enhance reproducibility of protocols, and advance toward standardizing the method*

## Conclusion

In conclusion, SRM represents a transformative advancement in the study of EVs, offering unparalleled resolution and precision that is crucial for single-particle analysis. By enabling detailed investigations into the morphology, molecular composition, and functional roles of EVs, SRM bridges critical gaps in our understanding of their biology and interactions within complex systems. However, as this review highlights, the adoption of SRM does not come without challenges, including the optimization of experimental protocols, the standardization of analysis workflows, and the careful selection of imaging reagents. Addressing them will be essential for leveraging the full potential of SRM in EV research.

The ongoing development of innovative methodologies, including those tailored to live-cell imaging and multiplexed analysis, holds promise for expanding the applications of SRM in this field. By fostering collaboration and standardization across research groups, the EV and SRM communities can pave the way for new discoveries, ultimately advancing both fundamental science and translational applications.

This review aims to inspire researchers to embrace the potential of SRM while equipping them with the foundational knowledge and practical tools to navigate its complexities effectively. To support this effort, we present a reporting template for the critical parameters of SRM applied to EVs, facilitating standardization, enhancing collaboration, and enabling meaningful comparisons across studies.

## Abbreviations
**Table 3: Acronyms and significations**

| Acronym | Signification | Acronym | Signification |
|---|---|---|---|
| 2D | Two-Dimensional | ISEV | International Society for Extracellular Vesicles |
| 3D | Three-Dimensional | LDL | Low-Density Lipoprotein |
| ADAM | a disintegrin and metalloproteinase | MBS | Maleimide-Activated Bovine Serum Albumin |
| ACE2 | Angiotensin-Converting Enzyme 2 | MEA | Microelectrode Array |
| AF4 | asymmetric flow field flow fractionation | NTA | Nanoparticle Tracking Analysis |
| AFM | Atomic Force Microscopy | OMV | Outer Membrane Vesicle |
| BME | Beta-Mercaptoethanol | PDAC | Pancreatic Ductal Adenocarcinoma |
| BSA | Bovine Serum Albumin | PALM | Photo-Activated Localization Microscopy |
| CA19-9 | Cancer Antigen 19-9 | PSF | Point Spread Function |
| CM dye | Cell Membrane Dye | PDL1 | Programmed Death-Ligand 1 |
| CLSM | Confocal Laser Scanning Microscopy | RTqPCR | Reverse Transcription Quantitative Polymerase Chain Reaction |
| DBSCAN | Density-Based Spatial Clustering of Applications with Noise | RNA | Ribonucleic Acid |
| DiD | 1,1'-Dioctadecyl-3,3,3',3'-Tetramethylindocarbocyanine Perchlorate | SEM | Scanning Electron Microscopy |
| DiO | 3,3'-Dihexyloxacarbocyanine Iodide | SARS-COV-19 | Severe Acute Respiratory Syndrome Coronavirus 2 |
| DNA | Deoxyribonucleic Acid | SMLM | Single-Molecule Localization Microscopy |
| DNAPAINT | DNA Points Accumulation for Imaging in Nanoscale Topography | SEC | size exclusion chromatography |
| EM | Electron Microscopy | SPIRIS | Spatially Resolved Infrared Spectroscopy |
| ESCRT | endosomal sorting complexes required for transport | STED | Stimulated Emission Depletion Microscopy |
| EGFR | Epidermal Growth Factor Receptor | SIM | Structured Illumination Microscopy |
| EpCAM | Epithelial Cell Adhesion Molecule | SRM | Super-Resolution Microscopy |
| EVs | Extracellular vesicles | SRSIM | Super-Resolution Structured Illumination Microscopy |
| GLOX | Glucose Oxidase | STORM/dSTORM | (direct) Stochastic Optical Reconstruction Microscopy |
| GPC1 | Glypican-1 | TIRF | Total Internal Reflection Fluorescence |
| GFP | Green Fluorescent Protein | TEM | Transmission Electron Microscopy |
| HSP | heat shock protein | TSG | Tumor Suppressor Gene |
| HBV | Hepatitis B Virus | TSG | Tumor susceptibility gene |
| HDL | High-Density Lipoprotein | TRPS | Tunable Resistive Pulse Sensing |
| HILO | High-Illumination Light Oblique Illumination | 2D | Two-Dimensional |
| HER2 | Human Epidermal Growth Factor Receptor 2 | VEGF | Vascular Endothelial Growth Factor |
| HLA-DR | Human Leukocyte Antigen-DR | VLDL | Very Low-Density Lipoprotein |
| HLA-G | Human Leukocyte Antigen-G | WGA | Wheat Germ Agglutinin |
| iSCAT | Interferometric Scattering Microscopy | | |

## Acknowledgments

This work was supported by the ANR EV fusion (ANR-21-CE11-0009). L. Alexandre is supported by the Programme d'Investissements d'Avenir launch by France and operated by ANR through the "Programmes et Equipements Prioritaires de Recherche (PEPR)" - CARN project (ANR-22-PEBI-0004). D. D'arrigo is supported by the Plan de reliance ANR Abbelight/MSC (CER PDR Abbelight-8255787). IVETh is supported by the IdEx Université Paris Cité, ANR-18-IDEX-0001, by the Region Ile de France under the convention SESAME 2019 – IVETh (EX047011) and via the DIM BioConvS, by the Région Ile de France and Banque pour l'Investissement (BPI) under the convention Accompagnement et transformation des filières projet de recherche et développement N° DOS0154423/00 & DOS0154424/00, DOS0154426/00 & DOS0154427/00, and Agence Nationale de la Recherche through the program France 2030 "Integrateur biotherapie-bioproduction" (ANR-22-AIBB-0002).

## Conflict of Interest

F.G and A.K.A.S. are cofounders and shareholders of the spin-off Evora Biosciences. Aditionally, A.K.A.S is a co-founder of the spin-off EverZom. F.G a,d A.K.A.S are also shareholders of the spin-off EverZom.